\documentclass[a4paper,11pt,oneside,openany]{article}

\usepackage{a4wide}

\usepackage{exscale}%
\usepackage{amsmath}%
\usepackage{amsfonts}%
\usepackage{times}
\usepackage{amsthm}
\usepackage{enumerate}
\usepackage{epsfig}
\newcommand{\la}{\langle} \newcommand{\ra}{\rangle}
\newcommand{\EE}{\mathbb{E}}

\newcommand{\RR}{\mathbb{R}}            
\newcommand{\NN}{\mathbb{N}}            
\newcommand{\CC}{\mathbb{C}}            

\newcommand{\ri}{\mathrm{ri}}

\newcommand{\Ran}{\mathop{\mathrm{Ran}}}     

\newcommand{\diag}{\mathop{\mathrm{diag}}}

\newcommand{\supp}{\mathop{\mathrm{supp}}}

\renewcommand{\thesection}
{\arabic{section}}                     
\renewcommand{\theequation}
{\thesection.\arabic{equation}}        

\newtheorem{theorem}{Theorem}[section]

\theoremstyle{definition}

\theoremstyle{remark}

\theoremstyle{plain}

\newcommand{\Mat}{\mathrm{M}}

\newcommand{\D}{\mathrm{d}}

\newcommand{\set}[2]{\{#1\, | \, #2\} }

\newcommand{\wick}[1]{\,:\!#1\!:\,}

\newcommand{\transpose}{\mathrm{t}}
\newcommand{\uone}{\underline{1}}

\usepackage{enumitem}
\setlist{itemsep=0pt}

\begin{document}
\bibliographystyle{amsplain}

\title{A Remark on Wick Ordering of Random Variables}
\author{ Jacob Schach M{\o}ller\\
Department of Mathematics, Aarhus University, Denmark}

\date{\today}

\maketitle

\abstract{This paper is a small note on the notation $\wick{q(X)}$, for the Wick ordering
of polynomials $q$ of random variables $X = (X_1,\dotsc,X_n)$, as introduced by Segal in \cite{Segal1969}. 
We argue that expressing $q(X)$ as another polynomial $p$ of a different set of random variables
$Y = (Y_1,\dotsc,Y_m)$,
does not give rise to a different Wick ordered random variable $\wick{p(Y)}$, provided the new random variables $Y_j$ are linear combinations of the $X_i$'s. 
}


\section{Introduction}

The notion of Wick ordering was introduced by Houriet and Kind \cite{HourietKind1949}, for bosonic field operators, and systematized and extended to mixed species of fields by Wick in \cite{Wick1950}. The topic of interest to us here is Wick ordering of polynomials of random variables, which in the case of Gaussian probability measures, corresponds to the purely bosonic setting through the identification of Segal's quantum fields with Gaussian random fields. See~\cite{DerezinskiGerard2013,Segal1969,Simon1974}. Wick ordered random variables also appears as a tool in probability theory \cite{AvramTaqqu1987,Vaiciulis2003}.

The notation commonly used in the literature for the Wick ordering of a polynomial $q(X)$ of a vector of random variables $X = (X_1,\dotsc,X_n)$ is $\wick{q(X)}$, a notation which goes back to Wick \cite{Wick1950} in the context of normal ordering polynomial expressions in annihilation and creation operators. However, the construction does not just depend on $q(X)$ as a random variable, but on $q$ as a polynomial in $n$ variables
and on the vector of random variables $X$ itself. That is, a different representation of the random variable
$q(X) = p(Y)$, will in general lead to a different Wick ordered random variable. However, it turns out that many 
natural representations of $q(X)$ does in fact produce the same
Wick ordered random variable.
The purpose of this note is to clarify the reason for this robustness.

We end the introduction with a brief overview of the paper.
In Section~\ref{Sec-Segal}, we recall the definition of some polynomials
associated with Borel probability measures on $\RR^n$.
They were introduced by Segal
in \cite{Segal1969}, and constitute a several variable version of Appell polynomials \cite{Appell1880}.
For brevity, we will refer to such polynomials as `Segal Polynomials', although (the more cumbersome) `Generalized Appell Polynomials' may be more appropriate.
We then turn to a transformation theorem for Segal polynomials, and we end Section~\ref{Sec-Segal} by
discussing some of its consequences. 
In Section~\ref{Sec-WickRV}, we recall and discuss Wick ordering of polynomials of random variables, the topic of our modest investigation. We shall see how the transformation theorem encodes the, perhaps surprising, robustness
in Wick's notation for Segal's random variable version of Wick ordering. Finally, in Section~\ref{Sec-Proof}, we have supplied a proof of the transformation theorem.

\section{Segal polynomials}\label{Sec-Segal}

Let $\mu$ be a Borel probability measure on $\RR^n$, and abbreviate 
$N = 1+\sup\set{k\in\NN_0}{\la |x|^k\ra_\mu<\infty}$, which is infinite if $\mu$ admits moments of all orders. 
Here $\la f\ra_\mu = \int_{\RR^n} f \,\D\mu$ denotes expectation w.r.t. the measure $\mu$, and $|\cdot|$ is 
the $1$-norm on $\RR^n$.

To such a measure, one can associate polynomials $p^\mu_\beta$ of order $|\beta|$, with $|\beta| <N$. Here $\beta\in \NN_0^n$ is a multi-index
and we write $\beta! = \beta_1!\cdots \beta_n!$ and $x^\beta = x_1^{\beta_1}\cdots x_n^{\beta_n}$, as usual.  
The polynomials are uniquely fixed by  the following two properties:
\begin{itemize}
\item $\forall\beta\in\NN_0^n$, with $|\beta|< N$, we have $\frac{\partial p^\mu_\beta}{\partial x_j}
= \beta_j p^\mu_{\beta-\delta_j}$.
\item $p^\mu_0 = 1$ and $\la p^\mu_\beta\ra_\mu = 0$, for all $\beta\in\NN_0^n$ with $0<|\beta|< N$.
\end{itemize}
Here $\delta_j$ is the multi-index with $(\delta_j)_i = \delta_{ij}$, the Kronecker delta.
The only term of maximal order 
$|\beta|$ in $p^\mu_\beta$ is $x^\beta$, which appears with coefficient equal to $1$.
That is, the polynomials $p_\alpha^\mu$ are monic.
In the special case of one dimension, $n=1$, the $p^\mu_k$'s are (up to normalization) Appell polynomials, cf.~\cite{Appell1880}. As mentioned in the introduction, we will refer to $p^\mu_\alpha$ as the Segal polynomials associated with the measure $\mu$.

 As a remark, we note that:  If $\mu$ is a Gaussian measure,
 then the Segal polynomials are orthogonal polynomials.
 If $\mu$ is Dirac's point measure, assigning the measure $1$ to $\{0\}$, then
  $p^\mu_\beta(x) = x^\beta$ for all $\beta$.
 If $\D\mu = (2\pi)^{-1/2}\exp(-x^2/2)\D x$, as a measure on $\RR$, then
  $p^\mu_k$ are the (monic) Hermite polynomials. 
 If the measure is a product measure, that is; $\mu = \mu_1\times\mu_2$ on $\RR^n = \RR^{n_1}\times\RR^{n_2}$,
 then $p^\mu_\beta(x) = p^{\mu_1}_{\beta_1}(x_1)  p^{\mu_2}_{\beta_2}(x_2)$,
 where $\beta = (\beta_1,\beta_2)\in\NN_0^{n_1}\times\NN_0^{n_2}$ and $x = (x_1,x_2)\in\RR^{n_1}\times\RR^{n_2}$. The last claim follows from uniqueness.
 Finally, the coefficients of the Segal polynomial $p^\mu_\beta$ are 
 polynomial expressions in expectation values $\la x^\alpha\ra_\mu$ with $\alpha\leq \beta$.

 Let $T\colon \RR^n\mapsto \RR^m$ be a linear transformation, i.e., an $m\times n$ real matrix
 with matrix elements $T_{i,j}$, where $i=1,\dotsc m$ and $j = 1,\dotsc ,n$.
The transformation $T$ induces a Borel probability measure $\mu_T$ on $\RR^m$, by the usual construction
$\mu_T(B) = \mu(T^{-1}(B))$, where $B$ denotes a Borel set in $\RR^m$ and $T^{-1}(B)$ its preimage in $\RR^n$ under $T$.  The transformed measure $\mu_T$ is supported on $\Ran(T)$, i.e. $\mu_T(C) = 0$, for any Borel set $C\subseteq\RR^m$ with $C\cap \Ran(T) = \emptyset$. 

 The transformed measure $\mu_T$ has moments of orders $k <N$, and hence has associated to it a family of polynomials $p^{\mu_T}_\alpha$, where $\alpha\in \NN_0^m$, with $|\alpha|<N$.
The result on which this note hinges, is a formula expressing $p^{\mu_T}_\alpha$, on the range of $T$,
in terms of $p^\mu_\beta$'s, with $|\beta| = |\alpha|$. 

In the following, $\Gamma\in\Mat_{m\times n}(\NN_0)$ is an $m\times n$ matrix (as is $T$) with entries
in $\NN_0$. The reader should think of $\Gamma$ as a multi-index with two labels.
We write $\Gamma! = \prod_{i,j}\Gamma_{i,j}!$ and $T^\Gamma= \prod_{i,j} T_{i,j}^{\Gamma_{i,j}}$, borrowing the standard multi-index notation. Finally, we abbreviate $\uone_\ell = (1,1,\dotsc,1)\in \RR^\ell$.
One should think of the rows and columns of $\Gamma$ as multi-indices and the usual multi-indices
$\Gamma\uone_n$ and $\Gamma^\transpose \uone_m$ now contains the lengths of the multi-indices sitting in the rows and columns of $\Gamma$, respectively. We are now ready to formulate:

\begin{theorem}\label{MainThm} Let $\mu$ be a Borel probability measure on $\RR^n$ and $T\colon \RR^n\mapsto\RR^m$ a linear transformation. Then for any $\alpha\in\NN_0^m$ and $x\in\RR^n$, we have
\begin{equation}\label{TransRulePols}
p^{\mu_T}_\alpha(Tx) = \sum_{\beta\in\NN_0^n, \ |\beta| = |\alpha|} A_{\alpha,\beta} \,p^\mu_\beta(x),
\end{equation}
where the transition coefficients $A_{\alpha,\beta}$ are given by the formula
\[
A_{\alpha,\beta} = \sum_{\stackrel{\Gamma\in\Mat_{m\times n}(\NN_0)}{\Gamma\uone_n = \alpha, \ \Gamma^\transpose \uone_m =\beta}}
\frac{\alpha!}{\Gamma!} \,  T^{\Gamma}.
\]
\end{theorem} 
 
 The formula in Theorem~\ref{MainThm} is more natural than it, perhaps, seems at a first glance, since 
\begin{equation}\label{deltacase}
(Tx)^\alpha = \sum_{\beta\in\NN_0^n,\, |\beta| = |\alpha|} A_{\alpha,\beta}\, x^\beta.
\end{equation}
This identity follows from the multinomial formula. In fact, it is also a special case of our theorem,
applied with $\mu$ equal to Dirac's point measure at zero, for which $\mu_T = \mu$ and $p^\mu_\beta(x) = x^\beta$.

 We end this section  by deriving some properties
 of Segal polynomials, which follow easily from Theorem~\ref{MainThm}. These properties may also be derived using the generating function
 \[
 G^\mu(\xi;x) = \sum_{\beta\in\NN_0^n} \frac{p^\mu_\beta(x)}{\beta!} \xi^\beta,
 \]
 viewed as a formal power series. Note that if the characteristic function
 $c_\mu(x) = \int_{\RR^n} \exp(-\ri x\cdot y)\,\D\mu(y)$ extends to an analytic function in a complex polydisc, around $0$, then the formal power series above converges absolutely on compact subsets of the same complex polydisc, around $\xi=0$, times $\CC^n$, as a function of $2n$ complex variables. Furthermore, $G^\mu(\xi;x) = \exp(x\cdot \xi)/c_\mu(\ri\xi)$. 
 See \cite{AvramTaqqu1987,Lukacs1970,Simon1974}. Here, however, we use instead our transformation theorem, which allows us to work at a fixed order of moments, thus giving shorter and conceptually simpler arguments. In fact, Theorem~\ref{MainThm} itself follows from expanding both sides
 of the identity $G^{\mu_T}(\xi;T x) = G^\mu(T^\transpose \xi;x)$ as power series
 in the variable $\xi$.
 Here $T^\transpose$ denotes the transpose of the matrix $T$. 
 The proof of Theorem~\ref{MainThm} we give in Section~\ref{Sec-Proof} does not rely on generating functions.
 
\medskip 
 
\noindent\textbf{Scaling:} Let $T=\diag\{c_1,\dotsc,c_n\}$ be a diagonal $n\times n$ matrix.
Then
$A_{\alpha,\beta}  = \delta_{\alpha\beta}c^\alpha$,
where we read $c = (c_1,\dotsc,c_n)$ as a vector to exploit the multi-index notation.
Hence
\begin{equation}\label{ScalingPolyForm}
p^{\mu_T}_\alpha(Tx) = c^\alpha p^\mu_\alpha(x). 
\end{equation} 
As a consequence, if $\mu$ is a reflection invariant measure, $\mu(-B) = \mu(B)$,
then $p^\mu_\alpha(-x) = (-1)^{|\alpha|} p^\mu_\alpha(x)$.
 
\medskip 
 
\noindent \textbf{Multinomial formula:} Let $T\colon \RR^n\mapsto \RR$ be given, i.e., $T$ acts by taking inner product
 with a vector $T = (c_1,\cdots,c_n)$.  Here the target space is one-dimensional, so
 the multi-index $\alpha$ is just a number $k\in\NN_0$. In this case
 $A_{k,\beta} = \frac{k!}{\beta!}\, c^\beta$ and,
 consequently,
 \begin{equation}\label{MultinomPolyForm}
 p^{\mu_T}_k(c_1x_1+\cdots + c_n x_n) = \sum_{\beta\in\NN_0^n,\,|\beta|=k} \frac{k!}{\beta!}\,  c^{\beta} p^\mu_\beta(x).
 \end{equation}
If $n=1$ as well, where $T\in\RR$, then $p^{\mu_T}_k(T x) = T^k p^\mu_k(x)$.   
 
\medskip

\noindent\textbf{Partial trace:} Let $x^\beta = x_1^{\beta_1}\cdots x_n^{\beta_n}$ be a monomial.
By partially tracing out variables, we may obtain $x^\beta$ from another monomial
$y^\alpha = y_1^{\alpha_1}\cdots y_m^{\alpha_m}$, if $m\in \NN$ and $\alpha\in\NN_0^m$ are such that we can choose a function
$J\colon \{1,\dotsc,m\}\to \{1,\dotsc,n\}$  with $\beta_j = \sum_{i : J(i) = j} \alpha_{i}$.
Then, by setting $y_i = x_{J(i)}$ we obtain that $x^\beta = y^\alpha$.

Let $\mu$ be a Borel probability measure on $\RR^n$.
We wish to study to what extend $p^\mu_\beta(x)$ can be obtained
from another Segal polynomial $p_\alpha^\nu(y)$ by partially tracing out variables  as above. We build a linear transformation
$T\colon \RR^n\mapsto \RR^m$ by setting $T e_{j} = \sum_{i: J(i) = j} e_i$. Note that 
$Tx = y$ with $y_i = x_{J(i)}$, $T^\transpose e_i = e_{J(i)}$ and, hence, $T^\transpose \alpha = \beta$.



With $T$ being the $m\times n$ matrix just introduced, we apply Theorem~\ref{MainThm} to compute $p_\alpha^{\mu_T}(Tx)$ for $\alpha\in \NN_0^m$ (provided moments of order $|\beta|$ exists).
Let $\alpha\in\NN_0^m$ and $\beta\in\NN_0^n$, with $|\alpha| = |\beta|$. To compute the transition coefficient $A_{\alpha,\beta}$, we let $\Gamma\in\Mat_{m\times n}(\NN_0)$
be such that $\Gamma\uone_n = \alpha$ and $\Gamma^\transpose \uone_m= \beta$.
For the product $T^{\Gamma}$
to be non-zero, we must have $\Gamma_{i,j} =0$, if $T_{i,j}=0$. That is,
for each $i$, we must require that $\Gamma_{i,j} = 0$, if $j\neq j(i)$, and $\Gamma_{i,j(i)} = \alpha_i$, in order to ensure that $\Gamma \uone_n = \alpha$. 
Since there is thus only one $\Gamma = \Gamma(\alpha)$ capable of contributing to the sum, we conclude that $A_{\alpha,\beta} = 0$, if
$\beta\neq \Gamma^\transpose \uone_m$.  Note that we can write this unique $\Gamma$ as $\Gamma= \diag\{\alpha_1,\alpha_2,\dotsc,\alpha_n\} T$, and hence,
$\beta = \Gamma^\transpose\uone_m = T^\transpose \alpha$ is the only multi-index for which $A_{\alpha,\beta}\neq 0$. And for this multi-index we have $T^\Gamma=1$.
We have now computed that
\[
A_{\alpha,T^\transpose \alpha} = 1 \quad \textup{and} \quad A_{\alpha,\beta}=0, \ \textup{if} \ \beta\neq T^\transpose \alpha.
\]

In conclusion, just as for the monomial $x^\beta$, the Segal polynomial $p_\beta^\mu(x)$ can be written as a partial trace of another Segal polynomial $p^{\nu}_\alpha(y)$, provided $T^\transpose \alpha = \beta$ and $\nu = \mu_T$. That is, $p^{\mu_T}_\alpha(Tx) = p^\mu_{T^\transpose \alpha}(x)$, for all $\alpha\in\NN_0^m$ (provided moments of order
$|\alpha|$ are defined). 
If $\mu$ is the point mass at zero, this just amounts to $(Tx)^\alpha = x^{T^\transpose \alpha}$, which was the monomial identity we started with.
 
\section{Wick ordered  random variables}\label{Sec-WickRV}
  Let $(\Omega,\Sigma, P)$ be a probability space. That is, $\Sigma$ is a $\sigma$-algebra of subsets of the set $\Omega$, and $P$ is a probability measure defined on $\Sigma$. By a random variable $X$, we understand as usual
  a $\Sigma$-measurable function $X\colon \Omega\mapsto \RR$. If we have several random variables, $X_1,\dotsc,X_n$, we form the vector $X = (X_1,\dotsc,X_n)$ as an $\RR^n$-valued random variable.
 
 We write $\EE$ for expectation with  respect to the probability measure $P$. 
 Given a random variable $X\colon \Omega\mapsto \RR^n$, 
 let $N = 1 + \sup\set{k\in\NN_0}{\EE[|X|^k]<\infty}$. That is, $X$ admits moments of order $k<N$.
  
  Given an $\RR^n$-valued random variable $X$, we can form an associated
 Borel probability measure $\mu_X$ on $\RR^n$, by setting $\mu_X(B) = P(X^{-1}(B))$.
 Here $X^{-1}(B)\in\Sigma$ denotes the preimage of the Borel set $B\subseteq \RR^n$.
 Note that if $X$ admit moments of orders $k<N$, then so does $\mu_X$.

 The Wick ordered monomials $\wick{X^\beta}$ are now defined for $\beta\in\NN_0^n$ with $|\beta|<N$,
 as the new (real-valued) random variable
 \[
 \wick{X^\beta} = p^{\mu_X}_\beta(X).
 \]
 This construction goes back to Segal \cite[Thm~1]{Segal1969}. See also the monographs \cite{DerezinskiGerard2013,Simon1974}. Let us pause to discuss Wick monomials, before we extend the notation to arbitrary polynomials.
 
 Below we consider linear transformations $Y = TX$ of vectors of random variables.
 The reason for the usefulness of Theorem~\ref{MainThm} is the transformation rule
 $(\mu_X)_T = \mu_Y$ for the associated Borel probability measures.
 Here $(\mu_X)_T(B) = \mu_X(T^{-1}(B))$ is the Borel measure on $\RR^m$ obtained by pushing forward  $\mu_X$.
 
 \medskip
 
 \noindent\textbf{Multinomial formula:}
 As an example of how one can use the transformation theorem to say something about Wick monomials,
 we derive the multinomial formula, hinted at in \cite{Simon1974}.
 Given a vector of real numbers $T=(c_1,\cdots,c_n)$, we can form a new random variable $Y= c_1 X_1+\cdots +c_n X_n\colon \Omega\mapsto \RR$. Since $\mu_Y = (\mu_X)_T$, we
 conclude from the multinomial formula \eqref{MultinomPolyForm} that
 \begin{equation}\label{MultiNomX}
  \wick{Y^k} = \sum_{\beta\in\NN_0^n, \, |\beta|=k}\frac{k!}{\beta!}\,c^\beta \wick{X^\beta},
 \end{equation}
 for any  $k$ with  $0\leq k<N$.
 This is the multinomial formula for Wick monomials.
 
 \medskip
 
 \noindent\textbf{Robustness of notation:}
 The notation $\wick{X^\beta}$ for Wick monomials suggests some robustness in how one chooses
 to represent the random variable $X^\beta$. This is in fact the issue, which led the author to study the properties of Segal polynomials to begin with.
 
 \noindent\emph{Scaling:} As a warm up, let us consider a simple question. Let $c =(c_1,\dotsc,c_n)$ be a vector of
 real numbers, such that $c^\beta = 1$, for some multi-index $\beta\in\NN_0^n$. 
 Putting $T = \diag\{c_1,\dotsc,c_n\}$ and $Y = TX$, we see that $X^\beta = Y^\beta$.
 The scaling law \eqref{ScalingPolyForm}, from the previous section, now tells us that
 \[
 \wick{Y^\beta} = p^{\mu_Y}_\beta(Y) = p^{(\mu_X)_T}_\beta(T X) = c^\beta p^{\mu_X}_\beta(X) = \wick{X^\beta}.
 \]
 This is a reassuring first sign of the notation $\wick{X^\beta}$ being a healthy choice.
 
 \noindent\emph{Partial trace:} Secondly, let us try to represent the random variable $X^\beta$ on a different form by
 repeating and/or removing copies of the $X_i$'s. To do this, we proceed as in the last paragraph of the previous section,
 and pick a matrix $T\colon\RR^n\mapsto\RR^m$, with exactly one non-zero entry in each row, 
 which should be equal to $1$. Then $Y=TX$ is a vector consisting of the same random variables, but with possible repetitions and/or omissions. Suppose $\alpha\in\NN_0^n$ is such that
 $\beta = T^\transpose\alpha\in\NN_0^m$. Then we have $X^\beta = Y^\alpha$, and we would like the two Wick ordered random variables, $\wick{X^\beta}$ and $\wick{Y^\alpha}$,  to coincide as well. 
 We compute, using the transformation rule from the last section,
 \[
 \wick{Y^\alpha} = p^{\mu_Y}_\alpha(Y) = p^{(\mu_X)_T}_\alpha(TX) = p^{\mu_X}_{T^\transpose \alpha}(X) = \wick{X^\beta}.
 \]
 That is, the Wick ordered monomials are not sensitive to how the monomial, into which the vector of random variables are inserted, has been represented.
 
 \medskip
 
 \noindent\textbf{A counter example:}
 While the above message suggests a robustness in how one reads $\wick{X^\beta}$, it is not correct
 that Wick ordering of any other representation of $X^\beta$, as a monomial, yields the same random variable.
 Here is an obvious counter example:
 Let $X$ be a random variable and define another random variable $Y = X^2$, such that
 $X^2 = Y^1$. We would like to compute $\wick{X^2}$ and $\wick{Y^1}$ and see if they are equal or not.
 (We assume of course that $X$ admits a second moment.)
 We have
 \begin{align*}
 \wick{X^2} & = p^{\mu_X}_2(X) = X^2- \la x\ra_{\mu_X} X + \la x\ra_{\mu_X}^2- \la x^2\ra_{\mu_X}\\
 & = X^2 -\EE[X] X + \EE[X]^2-\EE[X^2].
 \end{align*}
 On the other hand
 $\wick{Y^1} = p^{\mu_Y}_1(Y) = Y-\la y \ra_{\mu_Y} = Y- \EE[Y] = X^2-\EE[X^2]$.
 Hence, we observe that if $\EE[X]\neq 0$, then $\wick{X^2} \neq \wick{Y^1}$.

 
 \medskip
 
 \noindent\textbf{Extension by linearity:} We may extend the notation for Wick ordering
 from monomials to polynomials by linearity. To be precise, let $X = (X_1,\dotsc,X_n)$ be a vector of random variables and $q(x) = \sum_{\alpha\in\NN_0^n} c_\alpha x^\alpha$ with $c_\alpha\in\RR$, only finitely many of which are non-zero.
 Since the $x^\alpha$'s form a basis for the vector space of real polynomials, we get a well-defined map
 $\wick{\cdot}$ taking real polynomials into random variables
 \[
 \wick{q(X)} = \sum_{\alpha\in\NN_0^n} c_\alpha \wick{X^\alpha}.
 \] 
 
  The extension to polynomials, however, opens up a hornets nest of notational ambiguities. For example, in the multinomial formula, we computed the Wick ordering
 of $Y^k = (c_1 X_1+\cdots c_n X_n)^k$. We may also use the usual multinomial formula to write
 $Y^k$ as a polynomial and then use the extension by linearity above to assign a Wick ordered random variable to $Y^k$. By a stroke of good fortune, the end result \eqref{MultiNomX} is the same!
 
 More generally, suppose $q$ is a polynomial of $n$ variables and $p$ is a polynomial of $m$ variables with $n,m\in\NN$. Suppose one can express $q(x)$ as $p$ 
 evaluated at a linear combination of the variables $x_1,\dotsc,x_n$.
 That is, $q(x) = p(y)$ with $y = T x$ and $T\in \Mat_{m\times n}(\RR)$.
 Denote by $c_\beta^q$ and $c^p_\alpha$ the coefficients of the polynomials $q$ and $p$ respectively.
 
 Let $X = (X_1,\dotsc,X_n)$ be a vector of random variables and form a new 
 vector of random variables
 $Y = (Y_1,\dotsc,Y_m)$ by setting $Y = T X$. Then we may ask if the two Wick ordered  random variables $\wick{q(X)}$ and $\wick{p(Y)}$  -- as claimed in the abstract -- are identical. To explore this question we compute first using \eqref{deltacase}
 \begin{equation*}
 q(x) = p(Tx) = \sum_{\alpha} c^p_\alpha (Tx)^\alpha=
 \sum_{|\alpha|=|\beta|} c^p_\alpha A_{\alpha,\beta} x^\beta = \sum_{\beta}
 \Bigl(\sum_{\alpha: |\alpha|=|\beta|} c^p_\alpha A_{\alpha,\beta}\Bigr) x^\beta,
 \end{equation*}
 such that $c^q_\beta = \sum_{\alpha: |\alpha|=|\beta|} c^p_\alpha A_{\alpha,\beta}$.
 From this equation it now follows that
 \begin{align}\label{MainId}
 \nonumber\wick{q(X)} & = \sum_{\beta} c^q_\beta\wick{X^\beta}= \sum_{|\alpha|=|\beta|} c^p_\alpha \,A_{\alpha,\beta} \wick{x^\beta}
  =\sum_{|\alpha|=|\beta|} c^p_\alpha\, A_{\alpha,\beta}\, p^{\mu_X}_\beta(X)\\
 & = \sum_{\alpha} c^p_\alpha\, p^{\mu_Y}_\alpha(TX) = \sum_{\alpha} c^p_\alpha \wick{Y^\alpha}
    = \wick{p(Y)}.
 \end{align}
 
 
 \medskip
 
 \noindent\textbf{Random fields:} While the above identities may seem a little contrived, they take on more urgency when viewed through the lens of random fields. A random field is a map $\phi$ taking elements $v$
 of a real vector space $V$ into random variables $\phi(v)$ on a given probability space $(\Omega,\Sigma,P)$.
 The field should be linear, i.e.,
 \[
 \forall v,w\in V,\ a\in\RR:\qquad \phi(v+a w) = \phi(v) + a\phi(w), \quad P\textup{-a.e.}
 \]
 Due to linearity, it is precisely linear combinations of random variables that are the relevant
 transformations to consider. Let $v_1,\dotsc,v_n\in V$, and let $w_i= T_{i,1} v_1,+\cdots +T_{i,n} v_n$,
 where the $T_{i,j}$'s are real and $i=1,\dotsc,m$. For a real polynomial $p$ in $m$ variables,
 we may use linearity to write $p(\phi(w_1),\dotsc,\phi(w_m)) = q(\phi(v_1),\dotsc,\phi(v_n))$, $P$-a.e., for a real
 polynomial $q$ in $n$ variables. We can now conclude from \eqref{MainId} that 
 \begin{equation}\label{BasicSP-ID}
 \wick{p(\phi(w_1),\dotsc,\phi(w_m))}=
 \wick{q(\phi(v_1),\dotsc,\phi(v_n))}, \quad P\textup{-a.e}.
 \end{equation}
 
 \medskip

 \noindent\textbf{The Wiener process:}
 As a more interesting example, take the Gaussian random field arising from the 
 pointwise defined Wiener process $\{B(x)\}_{x\geq 0}$
 with $B(0)=0$, $\EE[B(x)] = 0$, $\EE[B(x)B(y)] = \min\{x,y\}$ and continuous sample paths. We may
 take $V$ to be
 the real vector space of continuous functions compactly supported in $\RR$.
 Then $\phi(f) = \int_0^\infty f(x) B(x) \,\D x$ for $f\in V$ (as a pointwise integral) is a (Gaussian) random field as considered above. 
 
 If $\supp f \subseteq (-\infty,b]$, we define approximants to $\varphi(f)$:
 \[
 Y_\ell(f) = \frac1{b\ell} \sum_{i=1}^\ell f( i b /\ell)B(i b /\ell)
 \]
  such that
 we have pointwise convergence $Y_\ell(f)\to \varphi(f)$ in $\Omega$.
 Suppose we have functions $f_1,\dotsc, f_n \in V$ given with $\supp f_j \subseteq (-\infty,b]$. Then we also have convergence of the expectations
 $\EE[Y_\ell(f_1)\cdots Y_\ell(f_n)]\to \EE[\varphi(f_1)\cdots \varphi(f_n)]$.
 To see this, we may assume that $n$ is even, since both sides vanish if $n$ is odd.
 For $n=2m$ the left hand side is a Riemann sum for the integral of the continuous compactly supported function $f_1(x_1)\cdots f_n(x_{2m})\EE[B(x_1)\cdots B(x_{2m})]$.
 That the expectation value is continuous follows from the identity
 \begin{equation*}
 \EE[B(x_1)\cdots B(x_{2m})] = \sum \EE[B(x_{i_1})B(x_{j_1})]\cdots \EE[B(x_{i_m})B(x_{j_m})],
 \end{equation*}
  where the sum is over all
 partitions of $1,\dotsc,2m$ into pairs $(i,j)$ with $i<j$. This is a special case of Wick's theorem for Gaussian Random Fields. It follows most easily from expanding generating functions \cite{Simon1974}.
 Put $Y_\ell = (Y_\ell(f_1),\dotsc,Y_\ell(f_n))$ and $Y = (\varphi(f_1),\dotsc,\varphi(f_n))$. Then for any multi-index $\beta$, we have $p^{\mu_{Y_\ell}}_\beta \to p^{\mu_Y}_\beta$. This follows since
 the coefficients in $p^{\mu_X}_\beta$ are polynomial expressions in $\la x^\alpha\ra_{\mu_X} = \EE[X^\alpha]$. Due to pointwise convergence of 
 $Y_\ell^\alpha\to Y^\alpha$ for all $\alpha$, we may finally conclude that $\wick{Y_\ell^\beta}\to \wick{Y^\beta}$ pointwise in $\Omega$ for any multi-index $\beta\in\NN_0^n$.
 
 Since the $Y_\ell(f_j)$'s are linear combinations of $B( i b /\ell)$, $i=1,\dotsc,\ell$, we may use \eqref{BasicSP-ID} to conclude that
 \[
 \wick{Y_\ell(f_1)\cdots Y_\ell(f_n)} = 
 \frac1{(b \ell)^{n}}\sum_{i_1,\dotsc,i_n=0}^\ell f_1(\tfrac{i_1 b}{\ell})\cdots f_n(\tfrac{i_n b }{\ell})
 \wick{B(\tfrac{i_1 b}{\ell})\cdots B(\tfrac{i_n b }{\ell})}\! .
 \]
 Observing that the the right-hand side is a Riemann sum (pointwise in $\Omega$), we may take $\ell$ to $\infty$ on both sides and arrive at the identity
 \begin{equation}
 \label{IntFormula}
 \begin{aligned}
 \wick{\phi(f_1)\cdots\phi(f_n)}
  & = \int_0^\infty\cdots \int_0^\infty f_1(x_1)\cdots f_n(x_n) \\
  & \qquad \times 
   \wick{B(x_1)\cdots B(x_n)} \D x_1\cdots \D x_n.
 \end{aligned}
 \end{equation}
 Note that $ (x_1,\dotsc,x_n)\mapsto \wick{B(x_1)\cdots B(x_n)}\!(\omega)$ is continuous
 for any $\omega\in\Omega$.
 
 We remark that there is nothing special about the Wiener process.
 Suppose we have random variables
 $\{\varphi(x)\}_{x\in\RR^d}$ indexed by points $x$ in $\RR^d$
 satisfying that $x\mapsto \varphi(x)(\omega)$ is, e.g.,  continuous for all $\omega\in\Omega$ and $(x_1,\dotsc,x_k)\mapsto \EE[\varphi(x_1)\cdots\varphi(x_k)]$ is continuous for all $k$ (supposing moments of order $k$ exists).
 The Wiener process (extended to zero on the negative half-axis) is an example.
 
 We can define a random field as above
 by setting $\varphi(f) = \int_{\RR^d} f(x) \varphi(x)\, \D x$ for $f\in V=C^\infty_\mathrm{c}(\RR^d;\RR)$. 
 Repeating the argument used for the Wiener Process, we see that \eqref{IntFormula} remains valid
 with $B(x_j)$ replaced by $\varphi(x_j)$ and the integration region
 replaced by $\RR^{d n}$.
 
 \medskip

\noindent\textbf{Conclusion:} The Segal polynomials $p^\mu_\beta$, by definition, carries many of the same 
combinatorial properties of ordinary polynomials. This manifests itself in Theorem~\ref{MainThm}, and is the reason that the notation used for 
Wick ordered polynomials of random variables has such surprising flexibility in interpretation.
Furthermore, the amount of robustness in the notation fits, hand in glove, with the linear structure of random fields.

 \section{Proof of the transformation theorem}\label{Sec-Proof}
 
 In this last section, we supply an elementary proof of Theorem~\ref{MainThm}. The proof goes by induction in the total order 
 $|\alpha|$. For $\alpha=0$ the identity reads $1=1$. 
Let $0 < N'<N$, and
assume that the theorem holds true for
multi-indices $\alpha\in\NN_0^n$, with $|\alpha| < N'$.

We begin by establishing an identity, relating the transition coefficients $A_{\alpha,\beta}$ at different order.
Let $1\leq \ell\leq n$, $\alpha\in \NN_0^n$ and $\beta\in\NN_0^m$, be such that
$|\alpha| = N'$ and $|\beta| = N'-1$. We write $E_{k\ell}$ for the $m\times n$ matrix unit,
with entries $(E_{k\ell})_{i,j} = \delta_{ki}\delta_{\ell j}$.
We have
\begin{align*}
& (\beta_\ell+1) A_{\alpha,\beta+\delta_\ell}  = 
\sum_{\stackrel{\Gamma\in \Mat_{m\times n}(\NN_0)}{\Gamma \uone_n= \alpha, 
\, \Gamma^\transpose\uone_m = \beta + \delta_\ell}} \frac{\bigl(\Gamma_{1,\ell} + \cdots + \Gamma_{m,\ell}\bigr)\alpha!}{\Gamma!} \, T^{\Gamma}\\
& \qquad = \  \sum_{k=1}^m \alpha_k T_{k,\ell}
\sum_{\stackrel{\Gamma\in \Mat_{m\times n}(\NN_0), \, \Gamma_{k,\ell}\geq 1}{\Gamma \uone_n= \alpha, 
\, \Gamma^\transpose\uone_m = \beta + \delta_\ell}}
 \frac{(\alpha-\delta_k)!}{(\Gamma-E_{k\ell})!} \, T^{\Gamma - E_{k\ell}}\\
& \qquad = \ \sum_{k=1}^m \alpha_k T_{k,\ell}
\sum_{\stackrel{\Gamma\in \Mat_{m\times n}(\NN_0)}
{\Gamma \uone_n = \alpha-\delta_k,\, \Gamma^\transpose \uone_m = \beta}} 
\frac{(\alpha-\delta_k)!}{\Gamma!} \, T^{\Gamma} = \ \sum_{k=1}^m \alpha_k T_{k,\ell}  A_{\alpha-\delta_k,\beta}.
\end{align*}
Note that only terms with $\alpha_k\geq 1$ contribute on the right-hand side.

Having established the above relation, we now
compute
\[
\frac{\partial p_\alpha^{\mu_T}\circ T}{\partial t_\ell}(t) = \sum_{k=1}^m T_{k,\ell} \frac{\partial p^{\mu_T}_\alpha}{\partial t_k}(T t) = \sum_{k=1}^m T_{k,\ell}\alpha_k p^{\mu_T}_{\alpha-\delta_k}(T t),
\]
and invoke the induction hypothesis to find that
\begin{align*}
\frac{\partial p_\alpha^{\mu_T}\circ T}{\partial t_\ell}(t)  = \! \sum_{\stackrel{\beta\in\NN_0^n}{|\beta|=N'-1}}\sum_{k=1}^m T_{k,\ell}
\alpha_k A_{\alpha-\delta_k,\beta} p^{\mu}_{\beta}(t) = \! \sum_{\stackrel{\beta\in\NN_0^n}{|\beta|=N'-1}} (\beta_\ell+1)A_{\alpha,\beta+\delta_\ell} p^{\mu}_{\beta}(t).
\end{align*}
On the other hand, we can compute the partial $t_\ell$-derivative of the right-hand side of the claimed equality \eqref{TransRulePols},
and get
\[
\sum_{\beta\in \NN_0^m, \, |\beta| = N'} A_{\alpha,\beta} \frac{\partial p^\mu_\beta}{\partial t_\ell}(t)
= \sum_{\beta\in \NN_0^m, \, |\beta| = N'} A_{\alpha,\beta} \beta_\ell p^\mu_{\beta-\delta_\ell}(t).
\]
A change of summation now establishes that the gradients of the left and right-hand side in 
\eqref{TransRulePols} coincide, and hence, 
we have established that the two polynomials coincide up to a constant factor.
Two such polynomials in $n$ variables are equal if and only if their expectation with respect to $\mu$
coincide.
But by construction of the polynomials, both sides of \eqref{TransRulePols} integrated with respect to $\mu$ gives zero, and we are done.

\noindent\textbf{Acknowledgment:} The author wishes to thank Volker Bach and Jan Derezi\'nski for helpful discussions. The first draft of this note was written during a stay at the Institute Henri Poincar\'e, April 2013, in connection with the trimester program ``Variational \& Spectral Methods in Quantum Mechanics".
The author thanks the organizers,  Maria Esteban and Mathieu Lewin, as well as  IHP, for hospitality. In fact, typing took place under the watchful eye of
Paul Appell, former rector of Universit\'e de Paris, whose portrait adorns a wall at IHP.


 \bibliographystyle{plain}

\providecommand{\bysame}{\leavevmode\hbox to3em{\hrulefill}\thinspace}
\providecommand{\MR}{\relax\ifhmode\unskip\space\fi MR }
\providecommand{\MRhref}[2]{%
  \href{http://www.ams.org/mathscinet-getitem?mr=#1}{#2}
}
\providecommand{\href}[2]{#2}

\end{document}